\documentclass[letterpaper]{article} 
\usepackage{aaai2026}  
\usepackage{times}  
\usepackage{helvet}  
\usepackage{courier}  
\usepackage[hyphens]{url}  
\usepackage{graphicx} 
\usepackage{natbib}  
\usepackage{caption} 
\usepackage{algorithm}
\usepackage{algorithmic}
\usepackage{booktabs}       
\usepackage{amsfonts}       
\usepackage{nicefrac}       
\usepackage{microtype}      
\usepackage{amsmath}
\usepackage{mathtools}
\usepackage{amsthm}
\usepackage{thm-restate}
\newtheorem{theorem}{Theorem}

\newtheorem{lemma}[theorem]{Lemma}
\newtheorem{corollary}[theorem]{Corollary}

\newtheorem{definition}[theorem]{Definition}

\usepackage{cleveref}
\usepackage{arydshln}
\usepackage{multicol, multirow}
\usepackage{paralist}

\newcommand{\R}{\mathbb{R}}
\newcommand{\E}{\mathbb{E}}

\newcommand{\len}{\mathrm{len}}

\usepackage{newfloat}
\usepackage{listings}
\DeclareCaptionStyle{ruled}{labelfont=normalfont,labelsep=colon,strut=off} 
\floatstyle{ruled}
\newfloat{listing}{tb}{lst}{}
\floatname{listing}{Listing}
\def\workdone{%
      \ifnum\value{eqfn}=0%
        \footnote{Work done while at Microsoft Research.}%
        \setcounter{eqfn}{\value{footnote}}%
      \else%
        \footnotemark[\value{eqfn}]%
      \fi%
    }%

\title{Incorporating Token Importance in Multi-Vector Retrieval}
\author{
    Archish S\thanks{Corresponding Author.}\workdone, Ankit Garg, Kirankumar Shiragur, Neeraj Kayal
}
\affiliations{
    Microsoft Research, India \\
    archish17@gmail.com,\{garga, kshiragur, neeraka\}@microsoft.com
}

\begin{document}

\maketitle

\begin{abstract}

ColBERT introduced a late interaction mechanism that independently encodes queries and documents using BERT, and computes similarity via fine-grained interactions over token-level vector representations. This design enables expressive matching while allowing efficient computation of scores, as the multi-vector document representations could be pre-computed offline. 
ColBERT models distance using a Chamfer-style function: for each query token, it selects the closest document token and sums these distances across all query tokens.

In our work, we explore enhancements to the Chamfer distance function by computing a weighted sum over query token contributions, where weights reflect the token importance. Empirically, we show that this simple extension, requiring only token-weight training while keeping the multi-vector representations fixed, further enhances the expressiveness of late interaction multi-vector mechanism. In particular, on the BEIR benchmark, our method achieves an average improvement of 1.28\% in Recall@10 in the zero-shot setting using IDF-based weights, and 3.66\% through few-shot fine-tuning.
\end{abstract}

\begin{links}
    \link{Code}{https://github.com/kayalneeraj/weighted_chamfer}
\end{links}

\section{Introduction}

Recent advances in natural language understanding have highlighted the power of deep language models for dense retrieval and document ranking. Building on this progress, \cite{khattab2020colbertefficienteffectivepassage} introduced ColBERT, a late interaction architecture that leverages BERT \cite{devlin2019bertpretrainingdeepbidirectional} to independently encode queries and documents. ColBERT enables efficient query-time scoring by precomputing token-level multi-vector representations for documents, while still capturing the rich semantics of deep models. ColBERT computes the relevance between a query and a document by comparing two sets of token-level vectors: for each query token, it identifies the most similar document vector using $\ell_2$ distance, and aggregates these distances across all query tokens. This approach has proven both efficient and effective \cite{khattab2020colbertefficienteffectivepassage}, outperforming earlier BERT-based and traditional retrieval models.

\begin{figure}
    \centering
    \includegraphics[width=0.98\linewidth]{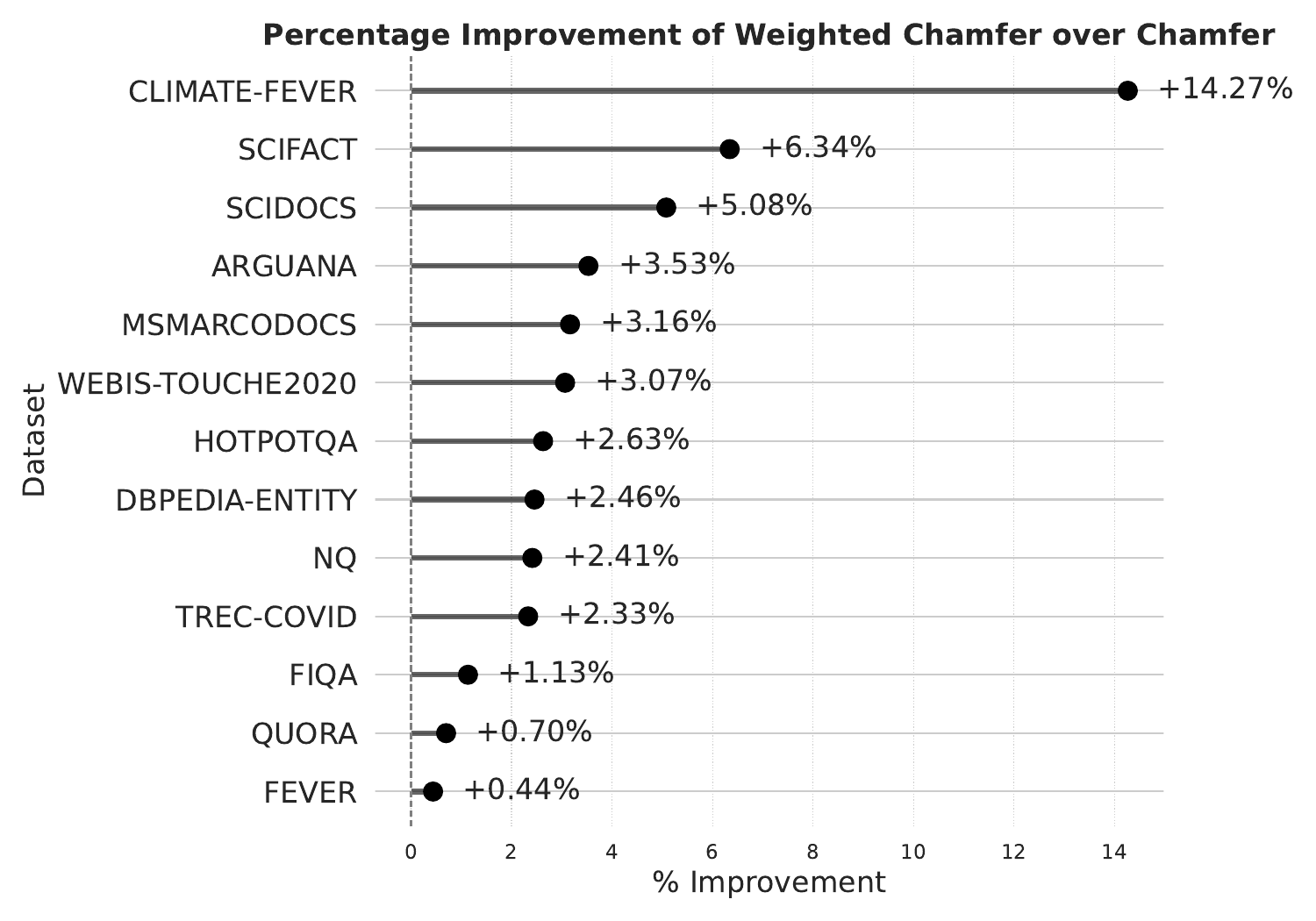}
    \caption{Effectiveness of Weighted Chamfer over Chamfer (ColBERTv2) on Recall@10 evaluated across the BEIR benchmark}
    \label{fig:enter-label}
\end{figure}

Traditional dense retrieval methods ~\cite{devlin2019bertpretrainingdeepbidirectional, nogueira2020passagererankingbert}, although popular for their ability to capture semantic information, yet often fall short in capturing fine-grained nuances, as they compress entire queries and documents into a single vector embedding. In contrast, classical retrieval systems like BM25~\cite{robertson1995okapi} excel at modeling term-level granularity through bag-of-words representations, though they lack semantic depth. Multi-vector retrieval systems aim to combine the strengths of both paradigms by representing each token with its own vector. However, current multi-vector approaches typically overlook term importance; a key factor that has contributed significantly to the success of BM25 and related models. 

In this work, we build upon the multi-vector retrieval framework and propose a novel extension to the Chamfer distance function. Rather than aggregating query token contributions uniformly, we introduce a weighted summation mechanism where each token’s contribution is scaled by a learned weight associated with its token ID. This weighing scheme captures the relative importance of tokens in determining query-document similarity, thereby enhancing the expressiveness of the model and improving ranking performance. This extension is particularly simple to incorporate into existing multi-vector pipelines and does not add any additional latency cost during the inference stage.

In ColBERT-style architectures, token vectors are typically normalized to unit norm prior to computing Chamfer distance~\cite{khattab2020colbertefficienteffectivepassage, santhanam2022colbertv2effectiveefficientretrieval}. Under this constraint, incorporating explicit token weights becomes especially important, as vector norms can no longer encode importance. While one might consider using unnormalized vectors to allow the model to implicitly encode token importance, e.g., by assigning lower norms to less relevant tokens, it remains unclear whether such behavior emerges reliably in practice. Although prior work has explored connections between embedding norms and word importance in static embeddings~\cite{oyama2022norm}, similar evidence for contextual embeddings remains limited.

One of the notable strengths of the ColBERT model is its impressive out-of-distribution generalization. Although trained solely on the MS MARCO dataset, ColBERT demonstrates strong zero-shot performance on the BEIR benchmark. In this work, we show that our simple weighted extension can give a significant improvement over ColBERT. In particular, a simple static weighting scheme based on inverse document frequency (IDF) computed from the document collection, yields an average gain of $1.28\%$ and maximum gain of $3.16\%$ in Recall@10 across BEIR datasets over ColBERTv2. Furthermore, by fine-tuning only the token weights (rather than the full model parameters) using limited human-labeled data, we observe an average gain of $3.66\%$ and maximum gain of $14.27\%$ over ColBERTv2. 

Our empirical results thus demonstrate the adaptability and effectiveness of our approach. These results underscore the potential of parameterized similarity functions in enhancing multi-vector retrieval, especially in zero-shot and low-resource settings. Given the limited availability of human-labeled relevance data and high cost in generating LLM based synthetic training data (an approach that has gained traction, e.g.,  \cite{dai2022promptagator, wang2023improving}), our method offers a lightweight alternative approach offering superior performance with limited resources. We also provide theoretical proofs under standard assumptions and techniques from statistical learning theory arguing about the sample complexity needed to learn the weights and also establish generalization bounds.

To learn the token weights, we employ a contrastive loss function, which tries to minimize the distance between the queries and positive documents relative to the negative ones. Notably, when the set of negative examples is fixed, the loss becomes a convex function with respect to the token weights, offering theoretical advantages and simplifying optimization. However, we find that training is most effective when hard negatives are selected iteratively, adapting to the current set of learned weights. A distinctive aspect of our training procedure is the use of a convex combination of two contrastive loss functions, each differing in the number of negatives per query, and we have found that this approach works best for finding the best set of token weights. This technique may prove valuable in other contrastive learning settings as well.

\section{Related Work}

Initial neural information retrieval (IR) approaches based on pre-trained transformers leverage BERT~\cite{devlin2019bertpretrainingdeepbidirectional, nogueira2020passagererankingbert, macavaney2019cedr} to model query-document relevance. These approaches also referred to as cross-encoders compute a similarity score by concatenating the query document pair and jointly encoding them using BERT. While effective, these methods are computationally expensive, as they require re-encoding each query-document pair during inference. Methods like SPLADE~\cite{formal2021spladesparselexicalexpansion, formal2021spladev2sparselexical} reformulate the sparse retrieval paradigm by explicitly invoking sparse regularization to improve efficiency.

In contrast to cross encoders, bi-encoders~\cite{karpukhin2020densepassageretrievalopendomain, macdonald2021single, qu2020rocketqa} encode queries and documents independently into high-dimensional vectors using separate BERT-based encoders and compute similarity through $\ell_2$ distance. This design allows for fast retrieval through pre-computed document embeddings and approximate nearest neighbor (ANN) search techniques ~\cite{johnson2019billion, malkov2018efficient, subramanya2019diskann}.

ColBERT~\cite{khattab2020colbertefficienteffectivepassage} addresses the lower expressivity of bi-encoders by representing queries and documents as sets of token-level embeddings and computing similarity using a late-interaction mechanism. This approach enables the pre-computation of document representations. Building on this, ColBERTv2~\cite{santhanam2022colbertv2effectiveefficientretrieval} introduces improved training strategies and indexing optimizations to support retrieval over billion-scale corpora. To further reduce latency, PLAID~\cite{santhanam2022plaidefficientenginelate} incorporates centroid interaction and pruning mechanisms, which eliminate low-scoring documents early in the search process, significantly improving efficiency without sacrificing retrieval quality.

MUVERA~\cite{dhulipala2024muveramultivectorretrievalfixed} seeks to bridge the gap between single-vector and multi-vector retrieval by approximating multi-vector similarity with fixed-dimension single vector encodings. While this design improves compatibility with highly optimized Maximum Inner Product Search (MIPS) solvers, it introduces a trade-off: its reliance on large high-dimensional vectors increases memory consumption, impacting storage efficiency.

More recent work such as XTR~\cite{lee2024rethinkingroletokenretrieval} refines ColBERT’s scoring by selecting and ranking only a subset of tokens through a modified training objective, thereby reducing computational cost. In parallel, ConstBERT~\cite{macavaney2025efficientconstantspacemultivectorretrieval} introduces a fixed-size representation for each document by pooling token embeddings into a smaller set of learned vectors, decoupled from individual input tokens. This improves storage efficiency and simplifies late interaction, making it more practical for downstream re-ranking. Other research on improving the latency and performance of multi-vector retrieval systems includes \cite{humeau2019poly, gao2021coil}.

\cite{qian2022multi} introduce an alignment framework to generalize ColBERT's late interaction mechanism. There are similarities to our approach but major differences in terms of parametrization and training objectives.

\section{Preliminaries}

While prior work on improving ColBERT has largely focused on reducing latency by pruning or compressing document vectors, our approach aims to enhance the expressiveness of the late interaction itself by reweighting query tokens. We employ the ColBERT model to embed the queries and the documents into multi-vector representations at the token-level. 

The ColBERT architecture comprises of \begin{inparaenum}[a)]
    \item query encoder 
    \item document encoder, and
    \item the late-interaction mechanism.
\end{inparaenum}
Let $q$ denote a query, which is a collection of tokens $\{q_i\}_{i=1}^{\len(q)}$ with $q_i \in [T]\, \forall\, i$, where $T$ denotes the total number of tokens in the universe. The ColBERT's query encoder encodes the query $q$ into a bag of fixed size unit-norm embeddings $E(q) = \{E_q(q_i)\}_{i=1}^{\len(q)} \subseteq \R^{d}$ while the document encoder encodes every document $d = (d_j)_{j=1}^{\mathrm{\len(d)}} $ in the document corpus $\mathcal{D}$ into another bag of unit-norm embeddings $E(d) = \{E_d(d_j)\}_{j=1}^{\len(d)}$ based on the tokens in $q$ or $d$ respectively. Throughout this work, the query and the document encoders are unchanged and used as is from ColBERTv2.

The late-interaction through Chamfer Distance (\cref{def:chamfer}) allows us for fast computation of relevance scores by enabling pre-computation of document token embeddings, while requiring only query token embeddings to be computed during the inference stage. The late-interaction mechanism involves matching every token $q_i$ in the query against the tokens in the document via the smallest $\ell_2$ distance between $E_q(q_i)$ and $E_d(d_j)$ vectors. We now formally define the Chamfer Distance.

\begin{definition}[Chamfer Distance~\cite{khattab2020colbertefficienteffectivepassage}]
    Given a query-document pair $(q,\, d)$, the Chamfer Distance is defined as 
    \begin{align}
        \label{def:chamfer}
        \mathrm{Chamfer} &= \eta(q,d) \coloneq  \\\nonumber
        \frac{1}{\len(q)} &\sum_{i=1}^{\len(q)} \underset{j \in [\len(d)]}{\min} \left\Vert E_q(q_i) - E_d(d_j) \right\Vert_2.
    \end{align}
\end{definition}

We replace the MaxSim operator introduced in ColBERT with the Euclidean distance instead to measure the document irrelevancy for elegance, aliased MinDist. This modification essentially keeps the performance unchanged. 

\section{Our Contribution}

In this work, we present Weighted Chamfer, a modified Chamfer Distance (\cref{def:wtdchamfer}) that improves performance on out-of-domain datasets, at virtually no increase in latency. This variant reflects the \textit{importance} of the query token $q_i$ in $q$, analogous to the word importance measures used in BM25 and related systems. The formal definition of Weighted Chamfer distance is as follows.

\begin{definition}[Weighted Chamfer Distance]
Given a query-document pair (q, d), the Weighted Chamfer distance is defined as 
\begin{align}
    \label{def:wtdchamfer}
    \mathrm{Weighted Chamfer} = \eta_{w}(q,d) \coloneq& \\\nonumber
    \frac{1}{\len(q)} \sum_{i=1}^{\len(q)} w_{q_i} \cdot \underset{j \in [\len(d)]}{\min}  &\left\Vert E_q(q_i) - E_d(d_j) \right\Vert_2
\end{align}
where $w_{q_i}$ are the weights associated with token $q_i$.
\end{definition}

\begin{figure}
    \centering
    \includegraphics[width=\linewidth]{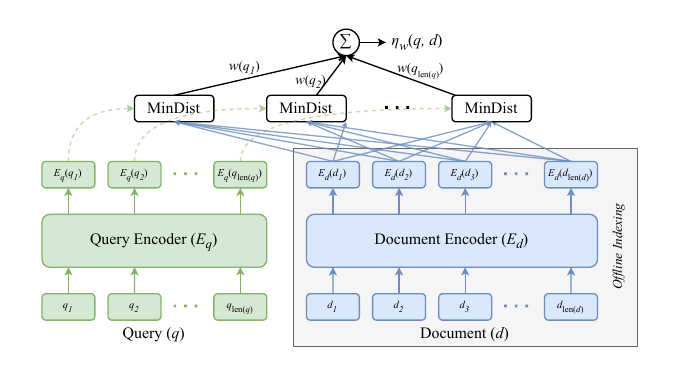}
    \caption{ColBERT Architecture with WeightedChamfer}
    \label{fig:pipeline}
\end{figure}

Note that Chamfer distance is a special case of Weighted Chamfer where all the token weights are set to $1$. 

We explore two different weighting techniques \begin{inparaenum}[a)]
    \item zero-shot setting (unsupervised) with no access to any relevance data\footnote{Relevance data consists of query-document pairs, each annotated with a label indicating their relevance.}, and
    \item few-shot setting (supervised) with limited access to relevance data.
\end{inparaenum}

\paragraph{Zero-shot Weighted Chamfer} 

In this setting, we choose the weights $w \in \R^{T}$ using statistics derived solely from the document corpus $\mathcal{D}$. Specifically, we use the widely used Inverse Document Frequency (IDF) as the weight for each token $t$. Formally, the IDF weight of a token $t$ is defined as:
\begin{equation}
    \text{IDF}(t) = \log \left( \frac{N - n(t) + 0.5}{n(t) + 0.5} + 1\right)
\end{equation}
where $N$ is the size of the corpus $\mathcal{D}$ and $n(t)$ is the number of documents that contain the token\footnote{For the tokens $t'$ that do not appear in the corpus (but seen in the queries, during inference), we assign weights to 0. For the special tokens (viz. \texttt{[MASK], [PAD], [QUERY], [DOCUMENT], [CLS]} tokens), we set weights to 0 or 1 based on the performance on the validation slice.} $t$. Since these token weights are computed solely from the token frequency in the document corpus, they do not require any labeled relevance data. Moreover, the computation of IDF weights has linear complexity in the total number of tokens in the corpus and can be further accelerated by sub-sampling documents, resulting in an approximate but efficient estimation of the weights.

\paragraph{Few-shot Weighted Chamfer}
In this setting, we prescribe a technique to learn the set of token weights using relevance data. We consider the widely used ranking loss, defined in~\cref{eq:rankingloss}, which tries to reduce the distance to the positive document $d^+ \in \mathcal{D}_q^+$ relative to the negative documents $\mathcal{D}_q^-$ for a query $q$. 

\begin{definition}[Cross-Entropy Loss]
    \label{def:rankingloss}
    Given a query $q$ and the set of positive (relevant) documents $\mathcal{D}_q^+ \subset \mathcal{D}$ and negative documents $\mathcal{D}_q^- \subseteq \mathcal{D} \backslash \mathcal{D}_q^+$ let
    \begin{align}
        \label{eq:rankingloss}
        \mathrm{CE}_{w}(q;\, \mathcal{D}_q^+, \mathcal{D}_q^-) &= \\ \nonumber
        -\sum\limits_{d \in \mathcal{D}_q^+} &\log\left( \frac{\exp\left( -\eta_{w}(q,d) \right)}{ \sum\limits_{d' \in \{\mathcal{D}_q^+ \cup \mathcal{D}_q^-\} } \exp\left( -\eta_{w}(q,d') \right)} \right).
    \end{align}
    The cross-entropy loss (ranking-loss) is defined as,
    $$\sum_q \mathrm{CE}_{w}(q;\, \mathcal{D}_q^+, \mathcal{D}_q^-)$$
\end{definition}
We remark here that the cross-entropy loss defined above is convex. We formally state this result next.
\begin{restatable}{lemma}{convexprogram}
    \label{lem:convex}
    For any query $q$ and sets $\mathcal{D}_q^+$ and $\mathcal{D}_q^-$, the function $\mathrm{CE}_{w}(q;\, \mathcal{D}_q^+, \mathcal{D}_q^-)$ is convex in variable $w$. Consequently, the ranking-loss function $\sum_q \mathrm{CE}_{w}(q;\, \mathcal{D}_q^+, \mathcal{D}_q^-)$ is also convex in $w$.
\end{restatable}

Since most datasets lack explicit annotations for irrelevant documents, the negative set is often very large. A common practice is to sample a subset of negatives (typically mined using a retriever) to make training computationally feasible. In our experiments, we found that combining losses (\cref{eq:overallloss}) computed over two sets of negatives, $\Lambda_q^{(1)} \subseteq \Lambda_q^{(2)} \subseteq \mathcal{D}_q^-$, leads to improved performance. We believe that this approach may also be beneficial in other contrastive learning applications.

\begin{align}
    \label{eq:overallloss}
         & \quad \quad \quad \quad \sum_{q} \mathcal{L}_{w}(q),  \text{ where }\\
    &\mathcal{L}_{w}(q) = \alpha \cdot \mathrm{CE}_{w}(q;\, \mathcal{D}_q^+, \Lambda_q^{(1)}) \nonumber \\ 
    & \quad \quad \quad \quad  + (1-\alpha) \cdot \mathrm{CE}_{w}(q;\, \mathcal{D}_q^+, \Lambda_q^{(2)}) \nonumber
\end{align}
By \Cref{lem:convex}, it follows immediately that the loss function defined above is convex. We formalize this result next.
\begin{corollary}
    For any query $q$ and sets $\mathcal{D}_q^+$, $\Lambda_q^{(1)}$ and $\Lambda_q^{(2)}$, the function $\mathcal{L}_{w}(q)$ is a convex in variable $w$. Consequently, the function $\sum_{q} \mathcal{L}_{w}(q)$ is also convex in $w$.
\end{corollary}

In our experiments, we find that instead of optimizing the loss function for a fixed set of negatives per query, choosing the closest (i.e., lowest-distance) negatives based on the current weights in an iterative fashion (to further encourage the model by forcing it to assign lower distances to positive documents by focusing on the hardest negatives) works best. Hence, in practice, we don't optimize the convex function $\sum_q \mathrm{CE}_{w}(q;\, \mathcal{D}_q^+, \mathcal{D}_q^-)$ but take gradient steps according to a series of convex functions which iteratively change according to the set of negatives. We provide complete details on our empirical performance in the results section.

In addition to our empirical results, we also provide theoretical bounds that characterize the limited amount of relevance data required to learn the weights. The theoretical models that we study don't exactly correspond to the empirical setting we are interested in. Instead they serve as reference points for understanding the sample complexity for learning token weights. Our theoretical contributions are twofold.

First, we study the problem of recovering the true hidden weights in a setting where query–document pairs are sampled i.i.d. from an underlying distribution, and their relevance score is given by the Weighted Chamfer score for some unknown true weights. In this setting, using the matrix Chernoff bound, we establish an upper bound on the sample complexity required to precisely recover the true hidden weights. This result is formal stated in the following theorem.

\begin{restatable}{theorem}{samplecomplexity}
    \label{thm:samplecomplexity}
    Given query-document pairs $(q,d)$ sampled i.i.d from a distribution $\mathfrak{D}$ and access to the weighted chamfer score from an underlying hidden weights. Let $A=\E_{(q,d) \sim \mathfrak{D}} \left[ \frac{1}{l} x_{(q,d)}x_{(q,d)}^{\top} \right]$, where $l = \len(q)$ and $x_{(q,d)}(t) = \underset{t' \in \len(d)}{\max} E_q(q_t)^{\top} E_d(d_{t'})$ if $t \in q$ and $0$ otherwise, then there exists an algorithm that takes 
    \begin{equation*}
        n \geq \Omega\left(\frac{1}{\lambda_{\min} (A)} \log \left(\frac{T}{\delta} \right)\right)
    \end{equation*}
    samples and with probability at least $1-\delta$, learns the true hidden weights.
\end{restatable}

Second, we relax the strong assumption of having access to Weighted Chamfer scores and instead assume access only to binary relevance labels for query–document pairs. We consider a hypothesis class consisting of threshold classifiers based on the Weighted Chamfer score, where the threshold $\tau$ is fixed and shared across all queries\footnote{In the ranking setting, each query could, in principle, have its own relevance threshold.}. Under this setting, we derive generalization bounds by reducing the problem of learning the weights to that of learning linear threshold functions. Using standard statistical learning theory techniques involving the VC dimension, we establish a bound on the sample complexity required to learn a Weighted Chamfer score–based classifier from this hypothesis class whose expected risk is within an additive $\epsilon$ of the optimum achievable by any function in the class. We defer the formal theoretical statements and their proofs to the appendix\footnote{Our theoretical results do not exactly model our empirical studies but instead provide a guiding direction for understanding their behavior.}. 

\section{Results}

We evaluate the performance of Weighted Chamfer in the re-ranking setting. For each query $q$, we employ BM25 (with Lucene parameters $k_1 = 1.5$ and $b = 0.75$) as the retriever, followed by Weighted Chamfer for re-ranking a small set of $k'$ documents ($k' = 1000$). Since $k'$ is small, we exhaustively score all the documents against the query $q$.

We use the ColBERTv2 checkpoint, trained on the MS MARCO Passage Ranking dataset \cite{bajaj2018msmarcohumangenerated}, with its parameters frozen for both zero-shot and few-shot evaluation settings. To assess performance, we evaluate the Weighted Chamfer and naive Chamfer\footnote{Note that ColBERTv2 uses Chamfer distance.} methods on a subset of the BEIR benchmark~\cite{thakur2021beirheterogenousbenchmarkzeroshot} (13 datasets not including MS MARCO), excluding datasets that are not publicly accessible. Re-ranking quality is primarily reported using Recall@10, MRR@10 and nDCG@10. Since ColBERTv2 is trained exclusively on MS MARCO, the results, on the other BEIR datasets, reflect out-of-domain performance.

\begin{table*}[!htp]
\centering
\renewcommand{\arraystretch}{1.5}
\Huge
\resizebox{\textwidth}{!}{%
\begin{tabular}{llccccccccccccc}
\toprule
\multicolumn{1}{c}{\multirow{3}{*}{\textbf{Task}}} & \multicolumn{1}{c}{\multirow{3}{*}{\textbf{Dataset}}}  &
\multicolumn{4}{c}{\textbf{Recall@10}} &
\multicolumn{4}{c}{\textbf{MRR@10}} &
\multicolumn{4}{c}{\textbf{nDCG@10}} \\
\cmidrule(lr){3-6} \cmidrule(lr){7-10} \cmidrule(lr){11-14}
& &
\multirow{2}{*}{\textbf{Chamfer}} & \multicolumn{2}{c}{\textbf{Weighted Chamfer}} & \multirow{2}{*}{\%$\Delta$} &
\multirow{2}{*}{\textbf{Chamfer}} & \multicolumn{2}{c}{\textbf{Weighted Chamfer}} & \multirow{2}{*}{\%$\Delta$}  &
\multirow{2}{*}{\textbf{Chamfer}} & \multicolumn{2}{c}{\textbf{Weighted Chamfer}} & \multirow{2}{*}{\%$\Delta$} \\
& & & Zero-Shot & Few-Shot & & & Zero-Shot & Few-Shot & & & Zero-Shot & Few-Shot & \\
\midrule

\multirow{1}{*}{\textbf{Passage Retrieval}} 
& MS MARCODOCS & 0.6984 & 0.7205 & 0.7205 & 3.16\% & 0.8926 & 0.9049 & 0.9049 & 1.38\% & 0.7285 & 0.7504 & 0.7504 & 3.01\% \\
\multicolumn{1}{r}{\textit{Average Performance}} & & & & & \textbf{+3.16\%} & & & & \textbf{+1.38\%} & & & & \textbf{+3.01\%} \\
\cdashline{1-15}[0.8pt/2pt]

\multirow{3}{*}{\textbf{Question Answering}} 
& NQ & 0.7622 & 0.7628 & 0.7806 & 2.41\% & 0.5296 & 0.5172 & 0.5351 & 1.04\% & 0.5707 & 0.5617 & 0.5787 & 1.40\% \\
& HOTPOTQA & 0.6963 & 0.6998 & 0.7146 & 2.63\% & 0.8922 & 0.8872 & 0.9019 & 1.09\% & 0.6954 & 0.6952 & 0.7113 & 2.29\% \\
& FIQA & 0.4412 & 0.4367 & 0.4462 & 1.13\% & 0.4334 & 0.4267 & 0.4375 & 0.95\% & 0.3630 & 0.3589 & 0.3672 & 1.16\% \\
\multicolumn{1}{r}{\textit{Avgerage Performance}} & & & & & \textbf{+2.06\%} & & & & \textbf{+1.03\%} & & & & \textbf{+1.62\%} \\
\cdashline{1-15}[0.8pt/2pt]

\multirow{1}{*}{\textbf{Entity Retrieval}} 
& DBPEDIA-ENTITY & 0.6716 & 0.6881 & 0.6881 & 2.46\% & 0.9320 & 0.9234 & 0.9245 & -0.80\% & 0.5417 & 0.5383 & 0.5407 & -0.18\% \\
\multicolumn{1}{r}{\textit{Avgerage Performance}} & & & & & \textbf{+2.46\%} & & & & \textbf{-0.80\%} & & & & \textbf{-0.18\%} \\
\cdashline{1-15}[0.8pt/2pt]

\multirow{2}{*}{\textbf{Argument Retrieval}} 
& ARGUANA & 0.6600 & 0.6600 & 0.6833 & 3.53\% & 0.2162 & 0.2131 & 0.2269 & 4.95\% & 0.3234 & 0.3206 & 0.3372 & 4.27\% \\
& WEBIS-TOUCHE2020 & 0.3327 & 0.3429 & 0.3429 & 3.07\% & 0.6174 & 0.5407 & 0.6173 & -0.02\% & 0.3591 & 0.3519 & 0.3577 & 0.39\% \\
\multicolumn{1}{r}{\textit{Avgerage Performance}} & & & & & \textbf{+3.29\%} & & & & \textbf{+2.46\%} & & & & \textbf{+2.33\%} \\
\cdashline{1-15}[0.8pt/2pt]

\multirow{1}{*}{\textbf{Duplicate Question Retrieval}} 
& QUORA & 0.9319 & 0.9384 & 0.9384 & 0.70\% & 0.8405 & 0.8502 & 0.8502 & 1.15\% & 0.8488 & 0.8576 & 0.8576 & 1.04\% \\
\multicolumn{1}{r}{\textit{Avgerage Performance}} & & & & & \textbf{+0.70\%} & & & & \textbf{+1.15\%} & & & & \textbf{+1.04\%} \\
\cdashline{1-15}[0.8pt/2pt]

\multirow{1}{*}{\textbf{Citation Prediction}} 
& SCIDOCS & 0.1575 & 0.1575 & 0.1655 & 5.08\% & 0.2860 & 0.3041 & 0.3041 & 6.33\% & 0.1558 & 0.1597 & 0.1635 & 4.94\% \\
\multicolumn{1}{r}{\textit{Avgerage Performance}} & & & & & \textbf{+5.08\%} & & & & \textbf{+6.33\%} & & & & \textbf{+4.94\%} \\
\cdashline{1-15}[0.8pt/2pt]

\multirow{3}{*}{\textbf{Fact Checking}} 
& FEVER & 0.9128 & 0.9128 & 0.9168 & 0.44\% & 0.8507 & 0.8466 & 0.8590 & 0.98\% & 0.8416 & 0.8389 & 0.8485 & 0.82\% \\
& CLIMATE-FEVER & 0.2804 & 0.2886 & 0.3204 & 14.27\% & 0.3044 & 0.2959 & 0.3385 & 11.20\% & 0.2391 & 0.2359 & 0.2675 & 11.88\% \\
& SCIFACT & 0.8136 & 0.8336 & 0.8652 & 6.34\% & 0.6714 & 0.6900 & 0.7194 & 7.15\% & 0.6994 & 0.7202 & 0.7485 & 7.02\% \\
\multicolumn{1}{r}{\textit{Avgerage Performance}} & & & & & \textbf{+7.02\%} & & & & \textbf{+6.44\%} & & & & \textbf{+6.57\%} \\
\cdashline{1-15}[0.8pt/2pt]

\multirow{1}{*}{\textbf{Bio-Medical IR}} 
& TREC-COVID & 0.9440 & 0.9660 & 0.9660 & 2.33\% & 0.9767 & 1.0000 & 1.0000 & 2.39\% & 0.7032 & 0.7169 & 0.7169 & 1.95\% \\
\multicolumn{1}{r}{\textit{Avgerage Performance}} & & & & & \textbf{+2.33\%} & & & & \textbf{+2.39\%} & & & & \textbf{+1.95\%} \\
\midrule

\multicolumn{1}{l}{\textbf{Average Performance}} & & & & & \textbf{+3.66\%} & & & & \textbf{+2.91\%} & & & & \textbf{+3.01\%} \\

\bottomrule

\end{tabular}
}
\caption{Recall@10, MRR@10, and nDCG@10 are reported after re-ranking top-1000 retrieved by BM25 on the BEIR benchmark using Chamfer and Weighted Chamfer (Zero-Shot and Few-Shot settings).}
\label{tab:metrics10_bm25retrieved}
\end{table*}

\begin{table*}[!htp]
\centering
\renewcommand{\arraystretch}{1.5}
\Huge
\resizebox{\textwidth}{!}{%
\begin{tabular}{llccccccccccccc}
\toprule
\multicolumn{1}{c}{\multirow{3}{*}{\textbf{Task}}} & \multicolumn{1}{c}{\multirow{3}{*}{\textbf{Dataset}}}  &
\multicolumn{4}{c}{\textbf{Recall@10}} &
\multicolumn{4}{c}{\textbf{MRR@10}} &
\multicolumn{4}{c}{\textbf{nDCG@10}} \\
\cmidrule(lr){3-6} \cmidrule(lr){7-10} \cmidrule(lr){11-14}
& &
\multirow{2}{*}{\textbf{Chamfer}} & \multicolumn{2}{c}{\textbf{Weighted Chamfer}} & \multirow{2}{*}{\%$\Delta$} &
\multirow{2}{*}{\textbf{Chamfer}} & \multicolumn{2}{c}{\textbf{Weighted Chamfer}} & \multirow{2}{*}{\%$\Delta$}  &
\multirow{2}{*}{\textbf{Chamfer}} & \multicolumn{2}{c}{\textbf{Weighted Chamfer}} & \multirow{2}{*}{\%$\Delta$} \\
& & & Zero-Shot & Few-Shot & & & Zero-Shot & Few-Shot & & & Zero-Shot & Few-Shot & \\
\midrule

\multirow{1}{*}{\textbf{Passage Retrieval}} 
& MS MARCODOCS & 0.3486 & 0.3512 & 0.345 & 0.75\% & 0.8992 & 0.8992 & 0.9054 & 0.69\% & 0.4217 & 0.4236 & 0.4292 & 1.78\% \\
\multicolumn{1}{r}{\textit{Avgerage Performance}} & & & & & \textbf{+0.75\%} & & & & \textbf{+0.69\%} & & & & \textbf{+1.78\%} \\
\cdashline{1-15}[0.8pt/2pt]

\multirow{3}{*}{\textbf{Question Answering}} 
& NQ & 0.7989 & 0.7861 & 0.7989 & 0.00\% & 0.546 & 0.5324 & 0.546 & 0.00\% & 0.5917 & 0.5782 & 0.5917 & 0.00\% \\
& HOTPOTQA & 0.6938 & 0.704 & 0.704 & 1.47\% & 0.8941 & 0.9009 & 0.9009 & 0.76\% & 0.6937 & 0.7043 & 0.7043 & 1.53\% \\
& FIQA & 0.4312 & 0.4283 & 0.4356 & 1.02\% & 0.4306 & 0.4232 & 0.4306 & 0.00\% & 0.3588 & 0.3551 & 0.3588 & 0.00\% \\
\multicolumn{1}{r}{\textit{Avgerage Performance}} & & & & & \textbf{+0.83\%} & & & & \textbf{+0.25\%} & & & & \textbf{+0.51\%} \\
\cdashline{1-15}[0.8pt/2pt]

\multirow{1}{*}{\textbf{Entity Retrieval}} 
& DBPEDIA-ENTITY & 0.6716 & 0.6881 & 0.6761 & 2.46\% & 0.932 & 0.9234 & 0.9245 & -0.80\% & 0.5417 & 0.5383 & 0.5407 & -0.18\% \\
\multicolumn{1}{r}{\textit{Avgerage Performance}} & & & & & \textbf{+2.46\%} & & & & \textbf{-0.80\%} & & & & \textbf{-0.18\%} \\
\cdashline{1-15}[0.8pt/2pt]

\multirow{2}{*}{\textbf{Argument Retrieval}} 
& ARGUANA & 0.6567 & 0.6567 & 0.6967 & 6.09\% & 0.2152 & 0.2112 & 0.2361 & 9.71\% & 0.3211 & 0.3183 & 0.3479 & 8.35\% \\ 
& WEBIS-TOUCHE2020 & 0.3306 & 0.3327 & 0.3327 & 0.64\% & 0.6163 & 0.585 & 0.6183 & 0.32\% & 0.357 & 0.3462 & 0.3573 & 0.08\% \\
\multicolumn{1}{r}{\textit{Avgerage Performance}} & & & & & \textbf{+3.36\%} & & & & \textbf{+5.02\%} & & & & \textbf{+4.22\%} \\
\cdashline{1-15}[0.8pt/2pt]

\multirow{1}{*}{\textbf{Duplicate Question Retrieval}} 
& QUORA & 0.9321 & 0.9381 & 0.9381 & 0.64\% & 0.8406 & 0.8504 & 0.8504 & 1.17\% & 0.8488 & 0.8576 & 0.8576 & 1.04\% \\
\multicolumn{1}{r}{\textit{Avgerage Performance}} & & & & & \textbf{+0.64\%} & & & & \textbf{+1.17\%} & & & & \textbf{+1.04\%} \\
\cdashline{1-15}[0.8pt/2pt]

\multirow{1}{*}{\textbf{Citation Prediction}} 
& SCIDOCS & 0.1575 & 0.1575 & 0.1655 & 5.08\% & 0.286 & 0.3041 & 0.2962 & 6.33\% & 0.1558 & 0.1597 & 0.1635 & 4.94\% \\
\multicolumn{1}{r}{\textit{Avgerage Performance}} & & & & & \textbf{+5.08\%} & & & & \textbf{+6.33\%} & & & & \textbf{+4.94\%} \\
\cdashline{1-15}[0.8pt/2pt]

\multirow{3}{*}{\textbf{Fact Checking}} 
& FEVER & 0.9142 & 0.912 & 0.9142 & 0.00\% & 0.8561 & 0.8515 & 0.8561 & 0.00\% & 0.8454 & 0.8418 & 0.8454 & 0.00\% \\
& CLIMATE-FEVER & 0.2728 & 0.2929 & 0.2929 & 7.37\% & 0.2945 & 0.289 & 0.2945 & 0.00\% & 0.2337 & 0.2328 & 0.2337 & 0.00\% \\
& SCIFACT & 0.8036 & 0.8202 & 0.8522 & 6.05\% & 0.6696 & 0.6849 & 0.7145 & 6.71\% & 0.6957 & 0.7133 & 0.7401 & 6.38\% \\
\multicolumn{1}{r}{\textit{Avgerage Performance}} & & & & & \textbf{+4.47\%} & & & & \textbf{+2.24\%} & & & & \textbf{+2.13\%} \\
\cdashline{1-15}[0.8pt/2pt]

\multirow{1}{*}{\textbf{Bio-Medical IR}} 
& TREC-COVID & 0.92 & 0.96 & 0.96 & 4.35\% & 0.98 & 1 & 1 & 2.04\% & 0.9272 & 0.9651 & 0.9651 & 4.09\% \\
\multicolumn{1}{r}{\textit{Avgerage Performance}} & & & & & \textbf{+4.35\%} & & & & \textbf{+2.04\%} & & & & \textbf{+4.09\%} \\
\midrule

\multicolumn{1}{l}{\textbf{Average Performance}} & & & & & \textbf{+3.04\%} & & & & \textbf{+2.12\%} & & & & \textbf{+2.24\%} \\

\bottomrule

\end{tabular}
}
\caption{Recall@10, MRR@10, and nDCG@10 are reported after end-to-end retrieval using ColBERTv2 on the BEIR benchmark using Chamfer and Weighted Chamfer (Zero-Shot and Few-Shot settings).}
\label{tab:metrics10_endtoend}
\end{table*}

We use BM25 as retriever due to its low latency and demonstrate the performance of Weighted Chamfer over Chamfer as a re-ranker. We believe the advantages presented here would generalize to other retrievers as well.

\paragraph{Weighted Chamfer} \Cref{tab:metrics10_bm25retrieved} and \Cref{tab:metrics10_endtoend} displays the Recall@10, MRR@10 and nDCG@10 of Chamfer and Weighted Chamfer post retrieval using BM25 and using the end-to-end ColBERTv2 pipeline respectively. In the zero-shot setting, the weights are computed solely from the document corpus and therefore do not require any relevance data. This is particularly useful in scenarios where human-labeled data is scarce or unavailable. Moreover, the static nature of these weights ensures low computational overhead. For datasets HOTPOTQA, MS MARCODOCS, SCIFACT, and WEBIS-TOUCHE2020, the reported results use zero weights for special tokens. For all the remaining datasets, a weight of 1 is assigned to special tokens (prior to normalization).

In the few-shot setting, with our prescribed training routine, we see a significant gain over the zero-shot setting, even when trained on a very small amount of relevance data\footnote{Note that TREC-COVID and WEBIS-TOUCHE2020 contain fewer than 50 queries.}. This demonstrates that Weighted Chamfer can effectively utilize limited supervision to substantially improve re-ranking quality. Moreover, since it maintains only a few additional parameters (per dataset) while sharing the underlying model, this approach is storage-friendly compared to maintaining multiple fine-tuned models. Our experiments suggest that Weighted Chamfer performs particularly well on datasets where the relevance signals are clear but differ significantly from those seen during training. This indicates that token weights are effective in adapting to dataset-specific importance.

\paragraph{Training Details} As discussed, we learn only the token weights, while keeping the underlying ColBERTv2 model frozen. We use the Adam optimizer~\cite{kingma2017adammethodstochasticoptimization}, and across all datasets, we use a learning rate of $10^{-4}$ and cosine scheduler to lower the learning rate to $10^{-8}$ over 100 iterations. The token weights are initialized uniformly, and after each iteration, we constrain their sum to $1$.

Due to the absence of predefined training splits\footnote{For datasets that include a train-dev split, we randomly sample from the training data to create train-validation subset. For the remaining datasets, we construct train-validation-dev splits by randomly sampling from the original dev set.} in some BEIR datasets, we adopt a random train–validation–test split to evaluate the effectiveness of our approach. We identify the optimal hyper-parameters, specifically, the sizes of the negative sets per query ($|\Lambda_1|$, $|\Lambda_2|$) and the interpolation factor $\alpha \in [0, 1]$ -- based on the performance on the validation split. The learned weights are then evaluated on the validation set, and we select the best-performing weights between IDF and learnt. If the learnt weights perform better, we retrain the model on the union of the training and validation sets to further enhance performance. We remark here that not all tokens are `seen' during the training procedure. To further improve performance, we assign IDF-based weights to tokens that are unseen in the training slice. While doing so, we ensure that the total weight assigned to seen tokens remains the same as in the IDF-weighted case. Essentially, the training procedure only re-weights the relative importance among the seen tokens, while assigning IDF-based weights to the unseen ones.

\Cref{tab:hyperparams} enumerates the choice of hyper-parameters used. Based on empirical results, when optimizing for Recall$@k$, we recommend setting the sizes of $\Lambda_1$ and $\Lambda_2$ to be slightly larger than $k$ and $10k$, respectively, and $\alpha=0.1$ in most cases. Through our experiments, we observe that datasets with limited supervision are particularly sensitive to the choice of hyper-parameters. Only in those cases, we do a hyper-parameter search over and choose the optimal values, but other wise resort to $|\Lambda_1| = 10$ and $|\Lambda_2| = 100$ with $\alpha=0.1$. 

\begin{table}[!htp]
\centering
\scriptsize
\begin{tabular}{lccc}
\toprule
\textbf{Dataset} & $\alpha$ & $|\Lambda_1|$ & $|\Lambda_2|$ \\
\midrule
ARGUANA & 0 & - & 500 \\
CLIMATE-FEVER & 0 & - & 1000 \\
DBPEDIA-ENTITY & 0.25 & 5 & 100 \\
FEVER & 0.1 & 10 & 100 \\
FIQA & 0.1 & 10 & 100 \\
HOTPOTQA & 0.1 & 100 & 1000\\
MSMARCODOCS & 0.1 & 100 & 1000  \\
NQ & 0.75 & 5 & 100 \\
QUORA & 0.1 & 10 & 100 \\
SCIDOCS & 0.5 & 100 & 250 \\
SCIFACT & 0.1 & 50 & 250\\
TREC-COVID & 0 & - & 100 \\
WEBIS-TOUCHE2020 & 0 & - & 100 \\
\bottomrule
\end{tabular}
\caption{Hyper-parameters used for Few-shot training of Weighted Chamfer}
\label{tab:hyperparams}
\end{table}

\section{Discussion}

In this work, we introduced Weighted Chamfer, an enhancement to the ColBERT re-ranking framework. By integrating token-level importance weights into the late-interaction mechanism, we improved retrieval performance without altering the core model architecture. 

In both zero-shot and few-shot settings, our proposed methods enhance re-ranking performance. In the zero-shot setting, weights are derived from corpus-level statistics without requiring any relevance supervision, while in the few-shot setting, our training routine learns optimal weights using limited relevance data. Empirically, our few-shot approach consistently outperforms the Chamfer baseline across all evaluation metrics; Recall@10, MRR@10, and nDCG@10. Importantly,  Weighted Chamfer is simple to deploy within existing re-ranking pipelines, making it particularly suitable for scenarios where relevance data is limited or supervision is costly.

Our findings indicate that modifying only the late-interaction mechanism can substantially enhance the expressiveness of the underlying embedding model, opening the door to more sophisticated distance functions. For instance, the distance function can be extended to promote matching between co-occurring query and document tokens, capturing multi-token phrases (e.g., ``white house'') while preserving lexical structure during retrieval. This direction is particularly promising, as it can be seamlessly integrated into existing re-ranking pipelines, even in low-supervision or resource-constrained scenarios.

\bibliography{weightedchamfer}

@article{oyama2022norm,
  title={Norm of word embedding encodes information gain},
  author={Oyama, Momose and Yokoi, Sho and Shimodaira, Hidetoshi},
  journal={arXiv preprint arXiv:2212.09663},
  year={2022}
}

@article{dai2022promptagator,
  title={Promptagator: Few-shot dense retrieval from 8 examples},
  author={Dai, Zhuyun and Zhao, Vincent Y and Ma, Ji and Luan, Yi and Ni, Jianmo and Lu, Jing and Bakalov, Anton and Guu, Kelvin and Hall, Keith B and Chang, Ming-Wei},
  journal={arXiv preprint arXiv:2209.11755},
  year={2022}
}

@article{wang2023improving,
  title={Improving text embeddings with large language models},
  author={Wang, Liang and Yang, Nan and Huang, Xiaolong and Yang, Linjun and Majumder, Rangan and Wei, Furu},
  journal={arXiv preprint arXiv:2401.00368},
  year={2023}
}

@misc{khattab2020colbertefficienteffectivepassage,
      title={ColBERT: Efficient and Effective Passage Search via Contextualized Late Interaction over BERT}, 
      author={Omar Khattab and Matei Zaharia},
      year={2020},
      eprint={2004.12832},
      archivePrefix={arXiv},
      primaryClass={cs.IR},
      url={https://arxiv.org/abs/2004.12832}, 
}

@misc{santhanam2022colbertv2effectiveefficientretrieval,
      title={ColBERTv2: Effective and Efficient Retrieval via Lightweight Late Interaction}, 
      author={Keshav Santhanam and Omar Khattab and Jon Saad-Falcon and Christopher Potts and Matei Zaharia},
      year={2022},
      eprint={2112.01488},
      archivePrefix={arXiv},
      primaryClass={cs.IR},
      url={https://arxiv.org/abs/2112.01488}, 
}

@misc{kingma2017adammethodstochasticoptimization,
      title={Adam: A Method for Stochastic Optimization}, 
      author={Diederik P. Kingma and Jimmy Ba},
      year={2017},
      eprint={1412.6980},
      archivePrefix={arXiv},
      primaryClass={cs.LG},
      url={https://arxiv.org/abs/1412.6980}, 
}

@inproceedings{robertson1995okapi,
author = {Robertson, Stephen and Walker, S. and Jones, S. and Hancock-Beaulieu, M. M. and Gatford, M.},
title = {Okapi at TREC-3},
booktitle = {Overview of the Third Text REtrieval Conference (TREC-3)},
year = {1995},
month = {January},
publisher = {Gaithersburg, MD: NIST},
url = {https://www.microsoft.com/en-us/research/publication/okapi-at-trec-3/},
pages = {109-126},
edition = {Overview of the Third Text REtrieval Conference (TREC–3)},
}

@misc{thakur2021beirheterogenousbenchmarkzeroshot,
      title={BEIR: A Heterogenous Benchmark for Zero-shot Evaluation of Information Retrieval Models}, 
      author={Nandan Thakur and Nils Reimers and Andreas Rücklé and Abhishek Srivastava and Iryna Gurevych},
      year={2021},
      eprint={2104.08663},
      archivePrefix={arXiv},
      primaryClass={cs.IR},
      url={https://arxiv.org/abs/2104.08663}, 
}

@misc{bajaj2018msmarcohumangenerated,
      title={MS MARCO: A Human Generated MAchine Reading COmprehension Dataset}, 
      author={Payal Bajaj and Daniel Campos and Nick Craswell and Li Deng and Jianfeng Gao and Xiaodong Liu and Rangan Majumder and Andrew McNamara and Bhaskar Mitra and Tri Nguyen and Mir Rosenberg and Xia Song and Alina Stoica and Saurabh Tiwary and Tong Wang},
      year={2018},
      eprint={1611.09268},
      archivePrefix={arXiv},
      primaryClass={cs.CL},
      url={https://arxiv.org/abs/1611.09268}, 
}

@misc{devlin2019bertpretrainingdeepbidirectional,
      title={BERT: Pre-training of Deep Bidirectional Transformers for Language Understanding}, 
      author={Jacob Devlin and Ming-Wei Chang and Kenton Lee and Kristina Toutanova},
      year={2019},
      eprint={1810.04805},
      archivePrefix={arXiv},
      primaryClass={cs.CL},
      url={https://arxiv.org/abs/1810.04805}, 
}

@misc{nogueira2020passagererankingbert,
      title={Passage Re-ranking with BERT}, 
      author={Rodrigo Nogueira and Kyunghyun Cho},
      year={2020},
      eprint={1901.04085},
      archivePrefix={arXiv},
      primaryClass={cs.IR},
      url={https://arxiv.org/abs/1901.04085}, 
}

@misc{karpukhin2020densepassageretrievalopendomain,
      title={Dense Passage Retrieval for Open-Domain Question Answering}, 
      author={Vladimir Karpukhin and Barlas Oğuz and Sewon Min and Patrick Lewis and Ledell Wu and Sergey Edunov and Danqi Chen and Wen-tau Yih},
      year={2020},
      eprint={2004.04906},
      archivePrefix={arXiv},
      primaryClass={cs.CL},
      url={https://arxiv.org/abs/2004.04906}, 
}

@inproceedings{subramanya2019diskann,
author = {Subramanya, Suhas Jayaram Subramanya and Devvrit  and Kadekodi, Rohan and Krishnaswamy, Ravishankar and Simhadri, Harsha},
title = {DiskANN: Fast Accurate Billion-point Nearest Neighbor Search on a Single Node},
booktitle = {NeurIPS 2019},
year = {2019},
month = {November},
url = {https://www.microsoft.com/en-us/research/publication/diskann-fast-accurate-billion-point-nearest-neighbor-search-on-a-single-node/},
}

@misc{santhanam2022plaidefficientenginelate,
      title={PLAID: An Efficient Engine for Late Interaction Retrieval}, 
      author={Keshav Santhanam and Omar Khattab and Christopher Potts and Matei Zaharia},
      year={2022},
      eprint={2205.09707},
      archivePrefix={arXiv},
      primaryClass={cs.IR},
      url={https://arxiv.org/abs/2205.09707}, 
}

@misc{dhulipala2024muveramultivectorretrievalfixed,
      title={MUVERA: Multi-Vector Retrieval via Fixed Dimensional Encodings}, 
      author={Laxman Dhulipala and Majid Hadian and Rajesh Jayaram and Jason Lee and Vahab Mirrokni},
      year={2024},
      eprint={2405.19504},
      archivePrefix={arXiv},
      primaryClass={cs.DS},
      url={https://arxiv.org/abs/2405.19504}, 
}

@misc{lee2024rethinkingroletokenretrieval,
      title={Rethinking the Role of Token Retrieval in Multi-Vector Retrieval}, 
      author={Jinhyuk Lee and Zhuyun Dai and Sai Meher Karthik Duddu and Tao Lei and Iftekhar Naim and Ming-Wei Chang and Vincent Y. Zhao},
      year={2024},
      eprint={2304.01982},
      archivePrefix={arXiv},
      primaryClass={cs.CL},
      url={https://arxiv.org/abs/2304.01982}, 
}

@misc{macavaney2025efficientconstantspacemultivectorretrieval,
      title={Efficient Constant-Space Multi-Vector Retrieval}, 
      author={Sean MacAvaney and Antonio Mallia and Nicola Tonellotto},
      year={2025},
      eprint={2504.01818},
      archivePrefix={arXiv},
      primaryClass={cs.IR},
      url={https://arxiv.org/abs/2504.01818}, 
}

@article{macdonald2021single,
  title={On single and multiple representations in dense passage retrieval},
  author={Macdonald, Craig and Tonellotto, Nicola and Ounis, Iadh},
  journal={arXiv preprint arXiv:2108.06279},
  year={2021}
}

@inproceedings{macavaney2019cedr,
  title={CEDR: Contextualized embeddings for document ranking},
  author={MacAvaney, Sean and Yates, Andrew and Cohan, Arman and Goharian, Nazli},
  booktitle={Proceedings of the 42nd international ACM SIGIR conference on research and development in information retrieval},
  pages={1101--1104},
  year={2019}
}

@article{qu2020rocketqa,
  title={RocketQA: An optimized training approach to dense passage retrieval for open-domain question answering},
  author={Qu, Yingqi and Ding, Yuchen and Liu, Jing and Liu, Kai and Ren, Ruiyang and Zhao, Wayne Xin and Dong, Daxiang and Wu, Hua and Wang, Haifeng},
  journal={arXiv preprint arXiv:2010.08191},
  year={2020}
}

@article{johnson2019billion,
  title={Billion-scale similarity search with GPUs},
  author={Johnson, Jeff and Douze, Matthijs and J{\'e}gou, Herv{\'e}},
  journal={IEEE Transactions on Big Data},
  volume={7},
  number={3},
  pages={535--547},
  year={2019},
  publisher={IEEE}
}

@article{malkov2018efficient,
  title={Efficient and robust approximate nearest neighbor search using hierarchical navigable small world graphs},
  author={Malkov, Yu A and Yashunin, Dmitry A},
  journal={IEEE transactions on pattern analysis and machine intelligence},
  volume={42},
  number={4},
  pages={824--836},
  year={2018},
  publisher={IEEE}
}

@article{gao2021coil,
  title={COIL: Revisit exact lexical match in information retrieval with contextualized inverted list},
  author={Gao, Luyu and Dai, Zhuyun and Callan, Jamie},
  journal={arXiv preprint arXiv:2104.07186},
  year={2021}
}

@article{humeau2019poly,
  title={Poly-encoders: Transformer architectures and pre-training strategies for fast and accurate multi-sentence scoring},
  author={Humeau, Samuel and Shuster, Kurt and Lachaux, Marie-Anne and Weston, Jason},
  journal={arXiv preprint arXiv:1905.01969},
  year={2019}
}

@misc{formal2021spladesparselexicalexpansion,
      title={SPLADE: Sparse Lexical and Expansion Model for First Stage Ranking}, 
      author={Thibault Formal and Benjamin Piwowarski and Stéphane Clinchant},
      year={2021},
      eprint={2107.05720},
      archivePrefix={arXiv},
      primaryClass={cs.IR},
      url={https://arxiv.org/abs/2107.05720}, 
}

@misc{formal2021spladev2sparselexical,
      title={SPLADE v2: Sparse Lexical and Expansion Model for Information Retrieval}, 
      author={Thibault Formal and Carlos Lassance and Benjamin Piwowarski and Stéphane Clinchant},
      year={2021},
      eprint={2109.10086},
      archivePrefix={arXiv},
      primaryClass={cs.IR},
      url={https://arxiv.org/abs/2109.10086}, 
}

@article{qian2022multi,
  title={Multi-vector retrieval as sparse alignment},
  author={Qian, Yujie and Lee, Jinhyuk and Duddu, Sai Meher Karthik and Dai, Zhuyun and Brahma, Siddhartha and Naim, Iftekhar and Lei, Tao and Zhao, Vincent Y},
  journal={arXiv preprint arXiv:2211.01267},
  year={2022}
}

@article{Tropp_2011,
   title={User-Friendly Tail Bounds for Sums of Random Matrices},
   volume={12},
   ISSN={1615-3383},
   url={http://dx.doi.org/10.1007/s10208-011-9099-z},
   DOI={10.1007/s10208-011-9099-z},
   number={4},
   journal={Foundations of Computational Mathematics},
   publisher={Springer Science and Business Media LLC},
   author={Tropp, Joel A.},
   year={2011},
   month=aug, pages={389–434} }

@article{vcdimensions,
author = {Vapnik, V. N. and Chervonenkis, A. Ya.},
title = {Necessary and Sufficient Conditions for the Uniform Convergence of Means to their Expectations},
journal = {Theory of Probability \& Its Applications},
volume = {26},
number = {3},
pages = {532-553},
year = {1982},
doi = {10.1137/1126059},
URL = { 
        https://doi.org/10.1137/1126059
},
eprint = { 
        https://doi.org/10.1137/1126059
}
}

\appendix
\section{Other Metrics}

\Cref{tab:metrics100_bm25retrieved} displays the Recall@100, MRR@100 and nDCG@100 re-ranking performance of the top-1000 documents retrieved using BM25 of Chamfer and Weighted Chamfer. \Cref{tab:metrics100_endtoend} displays the Recall@100, MRR@100 and nDCG@100 of Chamfer and Weighted Chamfer using the end-to-end ColBERTv2 pipeline.

\begin{table*}[!t]
\centering
\renewcommand{\arraystretch}{1.5}
\Huge
\resizebox{\textwidth}{!}{%
\begin{tabular}{llccccccccccccc}
\toprule
\multicolumn{1}{c}{\multirow{3}{*}{\textbf{Task}}} & \multicolumn{1}{c}{\multirow{3}{*}{\textbf{Dataset}}}  &
\multicolumn{4}{c}{\textbf{Recall@100}} &
\multicolumn{4}{c}{\textbf{MRR@100}} &
\multicolumn{4}{c}{\textbf{nDCG@100}} \\
\cmidrule(lr){3-6} \cmidrule(lr){7-10} \cmidrule(lr){11-14}
& &
\multirow{2}{*}{\textbf{Chamfer}} & \multicolumn{2}{c}{\textbf{Weighted Chamfer}} & \multirow{2}{*}{\%$\Delta$} &
\multirow{2}{*}{\textbf{Chamfer}} & \multicolumn{2}{c}{\textbf{Weighted Chamfer}} & \multirow{2}{*}{\%$\Delta$}  &
\multirow{2}{*}{\textbf{Chamfer}} & \multicolumn{2}{c}{\textbf{Weighted Chamfer}} & \multirow{2}{*}{\%$\Delta$} \\
& & & Zero-Shot & Few-Shot & & & Zero-Shot & Few-Shot & & & Zero-Shot & Few-Shot & \\
\midrule

\multirow{1}{*}{\textbf{Passage Retrieval}} 
& MS MARCODOCS & 0.3502 & 0.3563 & 0.3590 & 2.51\% & 0.8932 & 0.8992 & 0.9054 & 1.37\% & 0.4158 & 0.4236 & 0.4276 & 2.84\% \\
\multicolumn{1}{r}{\textit{Avgerage Performance}} & & & & & \textbf{+2.51\%} & & & & \textbf{+1.37\%} & & & & \textbf{+2.84\%} \\
\cdashline{1-15}[0.8pt/2pt]

\multirow{3}{*}{\textbf{Question Answering}} 
& NQ & 0.8906 & 0.8906 & 0.8906 & 0.00\% & 0.5355 & 0.5355 & 0.5394 & 0.73\% & 0.6008 & 0.6008 & 0.6038 & 0.50\% \\
& HOTPOTQA & 0.8014 & 0.8115 & 0.8143 & 1.61\% & 0.8933 & 0.8885 & 0.9030 & 1.09\% & 0.7224 & 0.7242 & 0.7370 & 2.02\% \\
& FIQA & 0.6515 & 0.6570 & 0.6570 & 0.84\% & 0.4414 & 0.4414 & 0.4444 & 0.68\% & 0.4211 & 0.4211 & 0.4236 & 0.59\% \\
\multicolumn{1}{r}{\textit{Avgerage Performance}} & & & & & \textbf{+0.82\%} & & & & \textbf{+0.83\%} & & & & \textbf{+1.04\%} \\
\cdashline{1-15}[0.8pt/2pt]

\multirow{1}{*}{\textbf{Entity Retrieval}} 
& DBPEDIA-ENTITY & 0.3697 & 0.3709 & 0.3709 & 0.32\% & 0.9320 & 0.9234 & 0.9245 & -0.80\% & 0.4276 & 0.4313 & 0.4313 & 0.87\% \\
\multicolumn{1}{r}{\textit{Avgerage Performance}} & & & & & \textbf{+0.32\%} & & & & \textbf{-0.80\%} & & & & \textbf{+0.87\%} \\
\cdashline{1-15}[0.8pt/2pt]

\multirow{2}{*}{\textbf{Argument Retrieval}} 
& ARGUANA & 0.9267 & 0.9233 & 0.9600 & 3.59\% & 0.2162 & 0.2265 & 0.2269 & 4.95\% & 0.3812 & 0.3781 & 0.4009 & 5.16\% \\
& WEBIS-TOUCHE2020 & 0.2617 & 0.2707 & 0.2744 & 4.85\% & 0.6228 & 0.6228 & 0.6228 & 0.00\% & 0.2727 & 0.2754 & 0.2764 & 1.36\% \\
\multicolumn{1}{r}{\textit{Avgerage Performance}} & & & & & \textbf{+4.22\%} & & & & \textbf{+2.48\%} & & & & \textbf{+3.26\%} \\
\cdashline{1-15}[0.8pt/2pt]

\multirow{1}{*}{\textbf{Duplicate Question Retrieval}} 
& QUORA & 0.9869 & 0.9872 & 0.9872 & 0.03\% & 0.8423 & 0.8518 & 0.8518 & 1.13\% & 0.8644 & 0.8719 & 0.8719 & 0.87\% \\
\multicolumn{1}{r}{\textit{Avgerage Performance}} & & & & & \textbf{+0.03\%} & & & & \textbf{+1.13\%} & & & & \textbf{+ 0.87\%} \\
\cdashline{1-15}[0.8pt/2pt]

\multirow{1}{*}{\textbf{Citation Prediction}} 
& SCIDOCS & 0.3485 & 0.3665 & 0.3665 & 5.16\% & 0.2938 & 0.3062 & 0.3117 & 6.09\% & 0.2190 & 0.2282 & 0.2293 & 4.70\% \\
\multicolumn{1}{r}{\textit{Avgerage Performance}} & & & & & \textbf{+5.16\%} & & & & \textbf{+6.09\%} & & & & \textbf{+4.70\%} \\
\cdashline{1-15}[0.8pt/2pt]

\multirow{3}{*}{\textbf{Fact Checking}} 
& FEVER & 0.9392 & 0.9397 & 0.9397 & 0.05\% & 0.8515 & 0.8474 & 0.8596 & 0.95\% & 0.8485 & 0.8459 & 0.8546 & 0.72\% \\
& CLIMATE-FEVER & 0.5037 & 0.5192 & 0.5192 & 3.08\% & 0.3044 & 0.3072 & 0.3495 & 14.82\% & 0.3012 & 0.2999 & 0.3252 & 7.97\% \\
& SCIFACT & 0.9220 & 0.9327 & 0.9327 & 1.16\% & 0.6755 & 0.6936 & 0.7247 & 7.28\% & 0.7236 & 0.7417 & 0.7659 & 5.85\% \\
\multicolumn{1}{r}{\textit{Avgerage Performance}} & & & & & \textbf{+1.43\%} & & & & \textbf{+7.68\%} & & & & \textbf{+4.85\%} \\
\cdashline{1-15}[0.8pt/2pt]

\multirow{1}{*}{\textbf{Bio-Medical IR}} 
& TREC-COVID & 0.7756 & 0.8134 & 0.8134 & 4.87\% & 0.9767 & 1.0000 & 1.0000 & 2.39\% & 0.5224 & 0.5432 & 0.5432 & 3.98\% \\
\multicolumn{1}{r}{\textit{Avgerage Performance}} & & & & & \textbf{+4.87\%} & & & & \textbf{+2.39\%} & & & & \textbf{+3.98\%} \\
\midrule

\multicolumn{1}{l}{\textbf{Avgerage Performance}} & & & & & \textbf{+2.16\%} & & & & \textbf{+3.13\%} & & & & \textbf{+2.88\%} \\

\bottomrule
\end{tabular}
}
\caption{Recall@100, MRR@100, and nDCG@100 are reported after re-ranking top-1000 retrieved by BM25 on the BEIR benchmark using Chamfer and Weighted Chamfer (Zero-Shot and Few-Shot settings).}
\label{tab:metrics100_bm25retrieved}
\end{table*}

\begin{table*}[!htp]
\centering
\renewcommand{\arraystretch}{1.5}
\Huge
\resizebox{\textwidth}{!}{%
\begin{tabular}{llccccccccccccc}
\toprule
\multicolumn{1}{c}{\multirow{3}{*}{\textbf{Task}}} & \multicolumn{1}{c}{\multirow{3}{*}{\textbf{Dataset}}}  &
\multicolumn{4}{c}{\textbf{Recall@100}} &
\multicolumn{4}{c}{\textbf{MRR@100}} &
\multicolumn{4}{c}{\textbf{nDCG@100}} \\
\cmidrule(lr){3-6} \cmidrule(lr){7-10} \cmidrule(lr){11-14}
& &
\multirow{2}{*}{\textbf{Chamfer}} & \multicolumn{2}{c}{\textbf{Weighted Chamfer}} & \multirow{2}{*}{\%$\Delta$} &
\multirow{2}{*}{\textbf{Chamfer}} & \multicolumn{2}{c}{\textbf{Weighted Chamfer}} & \multirow{2}{*}{\%$\Delta$}  &
\multirow{2}{*}{\textbf{Chamfer}} & \multicolumn{2}{c}{\textbf{Weighted Chamfer}} & \multirow{2}{*}{\%$\Delta$} \\
& & & Zero-Shot & Few-Shot & & & Zero-Shot & Few-Shot & & & Zero-Shot & Few-Shot & \\
\midrule

\multirow{1}{*}{\textbf{Passage Retrieval}} 
& MS MARCODOCS & 0.3486 & 0.3512 & 0.345 & 0.75\% & 0.8992 & 0.8992 & 0.9054 & 0.69\% & 0.4217 & 0.4236 & 0.4292 & 1.78\% \\
\multicolumn{1}{r}{\textit{Average Performance}} & & & & & \textbf{+0.75\%} & & & & \textbf{+0.69\%} & & & & \textbf{+1.78\%} \\
\cdashline{1-15}[0.8pt/2pt]

\multirow{3}{*}{\textbf{Question Answering}} 
& NQ & 0.9556 & 0.9461 & 0.9556 & 0.00\% & 0.5527 & 0.539 & 0.5527 & 0.00\% & 0.6276 & 0.6145 & 0.6276 & 0.00\% \\
& HOTPOTQA & 0.7886 & 0.7886 & 0.7886 & 0.00\% & 0.8952 & 0.9021 & 0.9021 & 0.77\% & 0.7184 & 0.7268 & 0.7268 & 1.17\% \\
& FIQA & 0.641 & 0.6402 & 0.641 & 0.00\% & 0.4385 & 0.4315 & 0.4385 & 0.00\% & 0.4176 & 0.413 & 0.4176 & 0.00\% \\
\multicolumn{1}{r}{\textit{Avgerage Performance}} & & & & & \textbf{+0.00\%} & & & & \textbf{+0.26\%} & & & & \textbf{+0.39\%} \\
\cdashline{1-15}[0.8pt/2pt]

\multirow{1}{*}{\textbf{Entity Retrieval}} 
& DBPEDIA-ENTITY & 0.3697 & 0.3709 & 0.3709 & 0.32\% & 0.932 & 0.9234 & 0.9245 & -0.80\% & 0.4276 & 0.4313 & 0.4313 & 0.87\% \\
\multicolumn{1}{r}{\textit{Avgerage Performance}} & & & & & \textbf{+2.46\%} & & & & \textbf{-0.80\%} & & & & \textbf{-0.18\%} \\
\cdashline{1-15}[0.8pt/2pt]

\multirow{2}{*}{\textbf{Argument Retrieval}} 
& ARGUANA & 0.9167 & 0.9133 & 0.95 & 3.63\% & 0.2269 & 0.2232 & 0.2496 & 10\% & 0.3776 & 0.3743 & 0.4058 & 7.47\% \\
& WEBIS-TOUCHE2020 & 0.256 & 0.2686 & 0.2686 & 4.92\% & 0.6216 & 0.5919 & 0.6236 & 0.32\% & 0.2675 & 0.2716 & 0.2716 & 1.53\% \\
\multicolumn{1}{r}{\textit{Avgerage Performance}} & & & & & \textbf{+4.28\%} & & & & \textbf{+5.16\%} & & & & \textbf{+4.50\%} \\
\cdashline{1-15}[0.8pt/2pt]

\multirow{1}{*}{\textbf{Duplicate Question Retrieval}} 
& QUORA & 0.9877 & 0.9899 & 0.9899 & 0.22\% & 0.8423 & 0.852 & 0.852 & 1.15\% & 0.8646 & 0.8725 & 0.8725 & 0.91\% \\
\multicolumn{1}{r}{\textit{Avgerage Performance}} & & & & & \textbf{+0.22\%} & & & & \textbf{+1.15\%} & & & & \textbf{+0.91\%} \\
\cdashline{1-15}[0.8pt/2pt]

\multirow{1}{*}{\textbf{Citation Prediction}} 
& SCIDOCS & 0.3485 & 0.3665 & 0.3665 & 5.16\% & 0.2938 & 0.3062 & 0.3117 & 6.09\% & 0.219 & 0.2282 & 0.2293 & 4.7\% \\
\multicolumn{1}{r}{\textit{Avgerage Performance}} & & & & & \textbf{+5.16\%} & & & & \textbf{+6.09\%} & & & & \textbf{+4.7\%} \\
\cdashline{1-15}[0.8pt/2pt]

\multirow{3}{*}{\textbf{Fact Checking}} 
& FEVER & 0.9389 & 0.9389 & 0.9389 & 0.00\% & 0.8568 & 0.8523 & 0.8568 & 0.00\% & 0.8518 & 0.8487 & 0.8518 & 0.00\% \\
& CLIMATE-FEVER & 0.4792 & 0.4917 & 0.4917 & 2.61\% & 0.3046 & 0.2994 & 0.3046 & 0.00\% & 0.2911 & 0.2901 & 0.2911 & 0.00\% \\
& SCIFACT & 0.9143 & 0.9277 & 0.9277 & 1.47\% & 0.6734 & 0.689 & 0.717 & 6.47\% & 0.7196 & 0.7364 & 0.7562 & 5.09\% \\
\multicolumn{1}{r}{\textit{Avgerage Performance}} & & & & & \textbf{+1.36\%} & & & & \textbf{+2.16\%} & & & & \textbf{+1.70\%} \\
\cdashline{1-15}[0.8pt/2pt]

\multirow{1}{*}{\textbf{Bio-Medical IR}} 
& TREC-COVID & 0.7238 & 0.7732 & 0.7732 & 6.83\% & 0.98 & 1 & 1 & 2.04\% & 0.7623 & 0.8093 & 0.8093 & 6.17\% \\
\multicolumn{1}{r}{\textit{Avgerage Performance}} & & & & & \textbf{+2.33\%} & & & & \textbf{+2.39\%} & & & & \textbf{+1.95\%} \\
\midrule

\multicolumn{1}{l}{\textbf{Average Performance}} & & & & & \textbf{+1.99\%} & & & & \textbf{+2.06\%} & & & & \textbf{+2.28\%} \\

\bottomrule

\end{tabular}
}
\caption{Recall@100, MRR@100, and nDCG@100 are reported after end-to-end retrieval using ColBERTv2 on the BEIR benchmark using Chamfer and Weighted Chamfer (Zero-Shot and Few-Shot settings).}
\label{tab:metrics100_endtoend}
\end{table*}

\section{Convexity of Ranking Loss}
In this section, we show the convexity of the cross entropy loss.

\convexprogram*

\begin{proof}
    \label{lemproof:convex}
    As sum of convex functions is convex, it is sufficient to argue that the function $\mathrm{CE}_{w}(q;\, \mathcal{D}_q^+, \mathcal{D}_q^-)$ is convex for every query $q$. Recall,
    \begin{align*}
        \mathrm{CE}_{w}(q;\, \mathcal{D}_q^+, \mathcal{D}_q^-) &= \\
        -\sum_{d \in \mathcal{D}_q^+} &\log\left( 
        \frac{\exp\left( -\eta_{w}(q,d) \right)}{\sum\limits_{d' \in \mathcal{D}_q^+ \cup \mathcal{D}_q^-} \exp\left( -\eta_{w}(q,d') \right)} 
        \right)
    \end{align*}
    where the scoring function $\eta_{w}(q,d)$ is defined as
    \begin{equation*}
        \eta_{w}(q,d) = \frac{1}{\len(q)} \sum_{i=1}^{\len(q)} w_{q_i} \cdot \min_{j \in [\len(d)]} \left\Vert E(q_i) - E(d_j) \right\Vert_2.
    \end{equation*}

    Note that $\eta_{w}(q,d)$ is a linear function in $w$ and we write $\eta_{w}(q,d) = \langle w, x_{(q, d)} \rangle$, where $x_{(q, d)} \in \mathbb{R}^T$ depends only on $q$ and $d$, and not on $w$. Rewriting the above defined loss for query $q$, we get
    \begin{align*}
        &\mathrm{CE}_{w}(q;\, \mathcal{D}_q^+, \mathcal{D}_q^-)
        = -\sum_{d \in \mathcal{D}_q^+} \log\Big( 
        \frac{e^{-\langle w, x_{(q, d)} \rangle}}{\sum\limits_{d' \in \mathcal{D}_q^+ \cup \mathcal{D}_q^-} e^{-\langle w, x_{(q, d')} \rangle}} 
        \Big) \\
        &= \sum_{d \in \mathcal{D}_q^+} \langle w, x_{(q, d)} \rangle + |\mathcal{D}_q^+| \cdot \log \Big( \sum\limits_{d' \in \mathcal{D}_q^+ \cup \mathcal{D}_q^-} e^{-\langle w, x_{(q, d')} \rangle} \Big)
    \end{align*}

    The second term is of the form,
    \begin{equation*}
        f(w) = \log \left( \sum_{i} e^{\langle w, x_i \rangle} \right),
    \end{equation*}
    where $x_i = -x_{(q, d')}$ for each $d' \in \mathcal{D}_q^+ \cup \mathcal{D}_q^-$. As the first term is linear in $w$, to prove convexity of the funtion $\mathrm{CE}_{w}(q;\, \mathcal{D}_q^+, \mathcal{D}_q^-)$ it is sufficient to show that $f$ is convex. Towards this, we compute the gradient of $f$:
    \begin{align*}
        \nabla f(w) & = \sum_i \frac{e^{\langle w, x_i \rangle}}{\sum_j e^{\langle w, x_j \rangle}} x_i \\ 
        & = \sum_i p_i x_i, \text{ where } p_i = \frac{e^{\langle w, x_i \rangle}}{\sum_j e^{\langle w, x_j \rangle}}.
    \end{align*}
    Furthermore, the Hessian of $f$ is as follows,
    \begin{align*}
        \nabla^2 f(w) 
        &= \frac{1}{\sum_j e^{\langle w, x_j \rangle}} 
        \sum_i e^{\langle w, x_i \rangle} x_i x_i^\top \\
        &- \left( \sum_i \frac{e^{\langle w, x_i \rangle}}{\sum_j e^{\langle w, x_j \rangle}} x_i \right) 
        \left( \sum_i \frac{e^{\langle w, x_i \rangle}}{\sum_j e^{\langle w, x_j \rangle}} x_i \right)^\top \\
        &= \sum_i p_i x_i x_i^\top - \left( \sum_i p_i x_i \right)\left( \sum_i p_i x_i \right)^\top.
    \end{align*}

    The expression is the covariance matrix of the $\{x_i\}$ under its softmax distribution $\{p_i\}$, and is thus positive semi-definite. Hence, $f(w)$ is convex and we conclude the proof.
\end{proof}

\section{Learning Weights}
We show the sample-complexity in learning the weights. As discussed earlier, given a query, document $(q, d)$ pair sampled from a distribution $\mathfrak{D}$; the Weighted Chamfer score can be written as $\langle w, x_{(q, d)} \rangle$, where both $w, x_{(q,d)} \in \R^{T}$. The $t^{\text{th}}$ coordinate of the vector $x_{(q,d)}$ is equal to $\underset{t' \in \len(d)}{\max} E_q(q_t)^{\top} E_d(d_{t'})$ if $t \in q$ and 0 otherwise. Using this representation and matrix Chernoff inequality (stated below) we prove our main theorem.

\begin{lemma}[Matrix Chernoff Inequality~\cite{Tropp_2011}]
    \label{lem:chernoff}
     For a finite sequence $S_k$ independent, random matrices with common dimension $T$ such that $\lambda_{\min} (S_k) > 0$ and $\lambda_{\max} (S_k) \leq L$, then for $S = \sum_k S_k$ we have
     \begin{equation*}
         \mathrm{Pr}\left[ \| \lambda_{\min}(S) \| \leq (1-\epsilon)\mu_{\min} \right] \leq T \left[ \frac{e^{-\epsilon}}{(1-\epsilon)^{(1-\epsilon)}} \right]^{\mu_{\min}/L}
     \end{equation*}
     for $\mu_{\min} = \lambda_{\min}\left( \E\left[ S\right] \right)$.
\end{lemma}

\samplecomplexity*
\begin{proof}

Let $X \in \R^{n \times T}$ be the data matrix, where each row corresponds to an i.i.d sample from the underlying distribution. The true weighted chamfer scores for each query document pair can be written as $s = X w_*$ for some unknown true weight vector $w_* \in \R^T$. We get to see $s$ and $X$ and our goal is to recover $w_*$. Note that the vector $w_*$ is uniquely identifiable if the matrix $X$ has a full column rank, i.e., $X^{\top} X$ is invertible. Note that, $X^{\top} X = \sum\limits_{(q,d) \sim \mathfrak{D}} x_{(q,d)}x_{(q,d)}^{\top}$.

Since $X$ consists of rows sampled randomly from the underlying distribution, for it to even have a chance of being invertible in the finite-sample regime, it is necessary that in the infinite-sample regime -- where $X$ converges to its expected matrix -- the expected matrix is invertible. The definition of expected matrix is as follows, 
\begin{equation*}
    A = \E_{(q,d) \sim \mathfrak{D}} \left[ \frac{1}{l} x_{(q,d)}x_{(q,d)}^{\top} \right]
\end{equation*}
In the remainder of the proof, we argue about the invertibility of $X$ by showing that after sufficient number of samples $X^{\top}X$ concentrates around its expected matrix $A$.  

Towards this, we define $y_{(q,d)} = \frac{1}{\sqrt{l}}A^{-1/2} x_{(q,d)}$, which follows $\E_{(q,d) \sim \mathfrak{D}} \left[ y_{(q,d)}y_{(q,d)}^{\top} \right] = I_T$. Instead of $X$, we equivalently work with matrix $Y$ and argue about its invertibility. We show that $Y^\top Y$ concentrates around identity matrix and thereby establishing its invertibility.

The norm of $x_{(q,d)}$ is bounded by the number of tokens in the query, hence 
\begin{equation*}
    \| y_{(q,d)} \|^2 = \frac{1}{l} x_{(q,d)}^\top A^{-1} x_{(q,d)} \leq \| A^{-1} \| \leq \frac{1}{\lambda_{\min} (A)} = L.
\end{equation*}

Applying Matrix Chernoff, we get, 
\begin{equation*}
     \mathrm{Pr}\left[ \| \lambda_{\min}(Y^\top Y) \| \leq (1-\epsilon)n \right] \leq T \left[ \frac{e^{-\epsilon}}{(1-\epsilon)^{(1-\epsilon)}} \right]^{n/L}.
\end{equation*}
We want with probability $1-\delta$ that $\lambda_{\min}(Y^\top Y)  \geq (1-\epsilon)n$, hence
\begin{align*}
    T \left[ \frac{e^{-\epsilon}}{(1-\epsilon)^{(1-\epsilon)}} \right]^{n/L} \leq \delta \\
    \implies n \geq \frac{1}{\lambda_{\min} (A) f(\epsilon)} \log \left(\frac{T}{\delta} \right)  
\end{align*}
where $f(\epsilon) = \epsilon - (1-\epsilon) \log(1-\epsilon)$. Fixing $\epsilon=1/2$, we get that for 
\begin{equation*}
    n \geq \Omega \left( \frac{1}{\lambda_{\min} (A)} \log \left(\frac{T}{\delta} \right) \right),
\end{equation*}
with probability at least $1-\delta$, the minimum eigenvalue of $Y^{\top}Y$ is at least $(1-\epsilon)n$, thus concluding the proof.
\end{proof}

\section{Generalization Bounds}

In this section, we study the generalization bounds for Weighted Chamfer given only the binary relevance labels. To establish the bound, we simplify the setting to a linear classification problem. Using standard tools from statistical learning theory, we characterize the sample complexity required to ensure that the empirical test loss closely approximates the true (expected) loss. 

Given samples $(x, y)$ drawn i.i.d. from a distribution $\mathfrak{D}$, the goal is to learn a hypothesis $h \in \mathcal{H}$ that minimizes a loss function $\ell$. We begin by defining the expected and empirical risks.

\begin{definition}[Expected Risk]
    The \emph{expected risk} (or test error) of a hypothesis $h \in \mathcal{H}$ is defined as
    \begin{equation*}
        L(h) \coloneq \mathbb{E}_{(x, y) \sim \mathfrak{D}} \left[ \ell\left((x, y), h\right) \right],
    \end{equation*}
    which measures the expected loss incurred by $h$ on a randomly sampled test example. The \emph{expected risk minimizer} $h^*$ is any hypothesis in $\mathcal{H}$ that minimizes this quantity
    \begin{equation*}
        h^* \in \underset{h \in \mathcal{H}}{\arg\min} \; L(h).
    \end{equation*}
\end{definition}

\begin{definition}[Empirical Risk]
    The \emph{empirical risk} (or training error) of a hypothesis $h \in \mathcal{H}$ on a dataset $\{(x^{(i)}, y^{(i)})\}_{i=1}^n$ drawn i.i.d. from $\mathfrak{D}$ is defined as
    \begin{equation*}
        \hat{L}(h) \coloneq \frac{1}{n} \sum_{i=1}^{n} \ell\left((x^{(i)}, y^{(i)}), h\right).
    \end{equation*}
    The \emph{empirical risk minimizer} $\hat{h}$ is any hypothesis that minimizes this empirical loss
    \begin{equation*}
        \hat{h} \in \underset{h \in \mathcal{H}}{\arg\min} \; \hat{L}(h).
    \end{equation*}
\end{definition}

We invoke the result of Uniform Convergence (\cref{lem:uniform}) to derive an upper bound on the sample complexity, specifically, the number of samples required to ensure that the empirical loss is within $\epsilon$ of the expected loss with high probability. Throughout this section, we consider loss functions that are $1$-Lipschitz.

\begin{lemma}[Uniform Convergence]
    \label{lem:uniform}
    Let $\mathcal{H}$ be a hypothesis class, and let $\ell: (\mathcal{X} \times \mathcal{Y}) \times \mathcal{H} \rightarrow \mathbb{R}$ be a loss function that is 1-Lipschitz. Suppose $\{(x^{(i)}, y^{(i)})\}_{i=1}^n$ are i.i.d. samples drawn from a distribution $\mathfrak{D}$ over $\mathcal{X} \times \mathcal{Y}$. Then, with probability at least $1 - \delta$, the following inequality holds
    \begin{align*}
       \left| L(\hat{h}) - L(h^*) \right| &\leq 4 \sqrt{\frac{2\, \mathrm{VC}(\mathcal{H}) \left( \log(n) + 1 \right)}{n}} \\ 
        &+ \sqrt{\frac{2 \log \left( \frac{2}{\delta} \right)}{n}},
    \end{align*}
    where $L(h)$ is the expected risk, $\hat{L}(h)$ is the empirical risk, and $\mathrm{VC}(\mathcal{H})$ denotes the VC dimension of the hypothesis class $\mathcal{H}$.
\end{lemma}

Using the uniform convergence result stated above we state and provide the proof for our main result of this section.

\begin{restatable}[Generalization Bound for Weighted Chamfer]{theorem}{generalizationbounds}
\label{thm:generalization-bound}

Let $(q,d)$ be query-document pairs sampled i.i.d. from a distribution $\mathfrak{D}$, with access to binary relevance labels for each pair. Consider the hypothesis class defined by the Weighted Chamfer score:
\begin{equation*}
\mathcal{H} = \left\{ (q,d) \mapsto \mathbb{I}\left[ \eta_w(q,d) \geq \tau \right] \mid w \in \mathbb{R}^T, \tau \in \mathbb{R} \right\}.
\end{equation*}

Then, for any $\epsilon > 0$ and $\delta \in (0,1)$, if
\begin{equation*}
    n \geq \Omega \left( \frac{T \log\left( \frac{1}{\epsilon} \right) + \log\left( \frac{1}{\delta} \right)}{\epsilon^2} \right),
\end{equation*}
with probability at least $1 - \delta$, the empirical risk minimizer $\hat{h} \in \mathcal{H}$ satisfies:
\begin{equation*}
    \left| L(\hat{h}) - L(h^*) \right| \leq \epsilon,
\end{equation*}
where $L(h)$ denotes the expected risk defined with respect to the ranking loss, and $h^*$ and $\hat{h}$ being the expected and empirical risk minimizers respectively.
\end{restatable}

\begin{proof}
    Consider the binary classification setting, where examples $(x_{(q,d)}, y)$ are drawn i.i.d. from an unknown distribution $\mathbb{R}^T \times \{0,1\}$. Each input $x_{(q,d)} \in \mathbb{R}^T$ represents a feature vector for a query-document pair, where $t^{\text{th}}$ coordinate of the vector $x_{(q,d)}$ is equal to $\underset{t' \in \len(d)}{\max} E_q(q_t)^{\top} E_d(d_{t'})$ if $t \in q$ and 0 otherwise, and $y \in \{0,1\}$ is the corresponding binary relevance label.

    Note that the Weighted Chamfer can be written as $\langle w, x_{(q, d)} \rangle$, and thus the hypothesis class reduces to the form
    \begin{equation*}
        \mathcal{H} = \left\{ x \mapsto \mathbb{I}\left[ w^\top x \geq \tau \right] \mid w \in \mathbb{R}^T, \tau \in \mathbb{R} \right\}.
    \end{equation*}
    
    This hypothesis class, in fact, corresponds to linear threshold classifiers in $\mathbb{R}^T$ (with a bias term $\tau$), it is well known that its VC dimension is bounded by $T + 1$~\cite{vcdimensions}, i.e., $\mathrm{VC}(\mathcal{H}) \leq T$.

     Directly from \Cref{lem:uniform}, it follows that for any $\epsilon > 0$ and $\delta \in (0, 1)$, with probability at least $1 - \delta$, the following holds
    \begin{equation*}
        \left| L(\hat{h}) - L(h^*) \right| \leq 
        4 \sqrt{\frac{2\, T \left( \log(n) + 1 \right)}{n}} + \sqrt{\frac{2 \log \left( \frac{2}{\delta} \right)}{n}}.
    \end{equation*}
    In particular, with $n$ training samples,
    \begin{equation*}
        4 \sqrt{\frac{2\, T \left( \log(n) + 1 \right)}{n}} + \sqrt{\frac{2 \log \left( \frac{2}{\delta} \right)}{n}} \leq \epsilon,
    \end{equation*}
    and since both terms are positive, it suffices that
    \begin{align*}
        4 \sqrt{\frac{2\, T \left( \log(n) + 1 \right)}{n}} &\leq \frac{\epsilon}{2} , \text{ and }\\
        \sqrt{\frac{2 \log \left( \frac{2}{\delta} \right)}{n}} &\leq \frac{\epsilon}{2}.
    \end{align*}
    Consequently, for
    \begin{equation*}
        n \geq \Omega \left( \frac{T \log\left( \frac{1}{\epsilon} \right) + \log\left( \frac{1}{\delta} \right)}{\epsilon^2} \right)
    \end{equation*}
    with probability at least $1 - \delta$, the empirical risk minimizer $\hat{h}$ satisfies
    \begin{equation*}
        \left| L(\hat{h}) - L(h^*) \right| \leq \epsilon.
    \end{equation*}
    We conclude the proof.
\end{proof}

\end{document}